\documentstyle{l-aa}

\def\mum1{\,\mu {\rm m}^{-1}}

\begin{document}

\thesaurus{	08(09.04.1;09.07.1;13.21.3)}  

\title{         Constraints on the properties of the
                $2175{\rm \AA}$ interstellar feature carrier}

\author{        F. Rouleau, Th. Henning \and R. Stognienko}


\offprints{     Th. Henning}

\institute{     Max-Planck-Gesellschaft, AG 
		``Staub in Sternentstehungsgebieten'', 
		Schillerg\"a\ss{}chen 2--3, D-07745 Jena, Germany}

\date{          Received 2 July 1996 / Accepted 15 November 1996}

\maketitle
\markboth{F. Rouleau et al.: Constraints on the properties of the
$2175{\rm \AA}$ interstellar feature carrier}{}

\begin{abstract}
Constraints on the possible shape and clustering, as well as optical 
properties, of grains responsible for the 2175 $\AA$ interstellar extinction 
feature (interstellar UV bump) are discussed. These constraints are based on 
the observation that the peak position of the interstellar UV feature is very 
stable (variations $\la$ 1\%), that the large variations in width ($\la$ 25\%) 
are uncorrelated with the peak position except for the widest bumps, and that 
the shape of the feature is described extremely well by a Drude profile. The UV 
extinction of small graphite grains is computed for various clustering models 
involving Rayleigh spheres. It is shown that compact clusters qualitatively 
satisfy the above observational constraints, except that the peak position falls 
at the wrong wavelength. As an alternative to graphite to model the optical 
properties of the interstellar UV feature carrier, a single-Lorentz oscillator 
model is considered, in conjunction with a clustering model based on clusters 
of spheres. Intrinsic changes in the peak position and width are attributed to 
variations in chemical composition of the grains, impacting upon the parameters 
of the Lorentz oscillator. Further broadening is attributed to clustering. These 
models are shown to satisfy the above observational constraints. Furthermore, 
the correlated shift of peak position with increased width, observed for the 
widest interstellar UV features, is reproduced. Models involving a second 
Lorentz oscillator to reproduce the FUV rise are also considered. The impact 
of this extra Lorentz oscillator on the peak position, width, and shape of the 
bump is investigated. Synthetic extinction curves are generated to model actual 
ones exhibiting a wide range of FUV curvatures. Physical mechanisms which might 
be of relevance to explain the variations of these optical properties are discussed.

\end{abstract}

\section{Introduction}

Ever since its discovery, the prominent $2175~{\rm \AA}$ UV feature
in the interstellar extinction curve has been attributed to graphite
--  or at least some form of carbon-rich material --
due to cosmic abundance constraints, and the fact that small graphite
particles produce a feature in roughly the right wavelength range.

The interstellar extinction curves along different lines of sight can
be reproduced amazingly well by the following simple parameterization
(Fitzpatrick \& Massa 1986; 1988; 1990)
\begin{eqnarray} \label{is_fit}
  k(x)&=&\frac{A(x)-A_{\rm V}}{A_{\rm B}-A_{\rm V}}\nonumber\\
      &=&a_1+a_2 x+\frac{a_3}{(x-x_{\rm max}^2/x)^2+\gamma^2}
+a_4 f_{\rm FUV}(x),\nonumber\\
\end{eqnarray}
where $x$ is the inverse wavelength (wavenumber in $\mu {\rm m}^{-1}$),
$A$ is the magnitude of extinction, $x_{\rm max}$ is close to the peak
wavenumber of the UV feature, and $a_1,\;a_2,\;a_3,\;a_4$ are
fitting parameters. The far-UV curvature is described by
$f_{\rm FUV}(x)=0.5392(x-5.9)^2+0.0564(x-5.9)^3$ for $5.9 < x < 8.0\mum1$ and
$f_{\rm FUV}(x)=0$ for $x < 5.9\mum1$. The function
$[(x-x_{\rm max}^2/x)^2+\gamma^2]^{-1}$ is
referred to as a ``Drude'' profile, though it is a profile produced
by small spheres described either by a single Lorentz or Drude
oscillator (Bohren \& Huffman 1983). This fit reproduces most if not
all interstellar extinction curves known to date, with no systematic
deviations or additional features or shoulders (at least, none
beyond the observational errors).

The main observational constraints concerning the interstellar UV feature 
(term with coefficient $a_3$ in Eq.~[\ref{is_fit}]) are
(Fitzpatrick \& Massa 1986; Jenniskens \& Greenberg 1993):
\begin{enumerate}
\item the remarkable constancy of its peak position,
$x_{\rm max}=4.60\pm 0.04\mum1$, though its small ($\la 1 \%$)
variations are larger than observational errors.
\item the wide range of variations in its width, $\gamma=1.0\pm 0.25\mum1$
(i.e. $\la 25 \%$).
\item the fact that the variations in peak position and
width are {\it uncorrelated},
except for the widest bumps, i.e. $\gamma \ga 1.2 \mum1$, for
which a systematic shift to {\it larger} $x$ is observed
(Cardelli \& Savage 1988; Cardelli \& Clayton 1991). The lines of
sight for which this is observed all pass through dark, dense regions.
\end{enumerate}

Various lines of sight show peculiar extinction curves depending
on the type of environment they pass through. For example, in the
hydrogen poor circumstellar environment of R~CrB, the bump is weaker
and shifted to $4.17 \mum1 (2400~{\rm \AA})$. Around carbon-rich
(and hydrogen-rich) asymptotic giant branch stars, the bump is
considerably weakened or absent (e.g., Snow et al.\ 1987).
In H\,{\sc ii} regions, the relative strength of the bump (parameter $a_3$)
is weaker than in most other environments (Jenniskens \& Greenberg 1993).
However, the range of bump widths is similar to that of the diffuse
interstellar medium. Broader bumps in H\,{\sc ii} regions
have their peak position shifted to {\it smaller} $x$.
In other dense environments, like
``bubbles'' (regions showing loops and filaments characteristic
of material swept up by stellar winds of OB stars in regions of
recent massive star formation), the strength of the bump is similar
to that of the diffuse ISM and no correlation is observed between
its width and its peak position.

In this paper constraints on the optical properties of the
purported interstellar UV feature carrier are derived considering
rather general arguments related to chemical composition (as modelled
by graphite or Lorentz oscillator models) and clustering (based on
direct computations and interpretation using a spectral representation).
In Sect.~\ref{theor_models} some current theoretical models attempting
to explain the characteristics of the interstellar UV feature are
discussed. Their main shortcomings are emphasized.
In Sect.~\ref{Graphite} the profile of the UV feature of graphite
is computed for various arrangements of touching spheres. The results
are compared qualitatively to the observational constraints. Apart
from the peak position falling at the wrong wavelength, the qualitative
agreement is found to be good for compact clusters. But variations
in chemical composition cannot be included directly in such a model.
Therefore, in Sect.~\ref{Single_Lorentz} a series of single-Lorentz
oscillator models are used in conjunction with clustering to
investigate the range of parameters (which simulate variations in
chemical composition) consistent with the observational
constraints. The clustering is modelled via a spectral representation
formalism. In Sect.~\ref{Two_Lorentz} the effects of adding a second
Lorentz oscillator to explain the FUV rise are discussed. Interstellar
curves along specific lines of sight exhibiting a wide range of FUV
curvatures are modelled and plausible (though non-unique) optical
constants for the bump grains along these lines of sight are derived.
In Sect.~\ref{Mechanisms} physical mechanisms relevant to the
expected optical properties of the UV feature carrier are discussed,
in particular, dehydrogenation and UV processing.

\section{Current theoretical models} \label{theor_models}
Detailed theoretical comparisons between electromagnetic scattering
models involving small graphite
grains and the interstellar UV feature are complicated
by the fact that graphite is a highly anisotropic material. In particular,
an approximation must be used when computations involving Mie theory 
(valid only for isotropic spheres) are carried out.
The MRN model (Mathis et al.\ 1977) and its variants,
using a size distribution of spherical grains of silicate
and graphite, can explain the mean
extinction curve, but fail to satisfy the above observational constraints.
Draine (1988) has studied the UV feature produced by
graphite particles using the discrete dipole approximation (DDA).
This method is ideally suited for handling the anisotropic
dielectric tensor of graphite.
He found that only particles of small elongation 
with equivalent radii in the range 100--200~${\rm \AA}$ could provide
a reasonable fit to typical interstellar UV features.

Draine \& Malhotra (1993) studied the variations in peak position, width,
and strength of the bump for various models based on a size distribution of
graphite grains to see whether they were compatible with the 
observational constraints. DDA calculations included spherical graphite
grains with an ice coating, spheroidal graphite grains,
and graphite spheres in contact with silicate spheres. All models using
variations in shape or coating produced correlations between the
peak position and width. Therefore, they concluded that the variations observed
must be due to changes in the dielectric properties of the grains
either through impurities or surface effects, rather than purely
``geometric'' effects.

Mathis (1994) considered a model consisting of a graphite oblate
spheroidal core and a coating of material represented by an
appropriately chosen single Lorentz oscillator.
The weakness of this model lies in the fact that
the shape of the grain had to be unreasonably fine-tuned in order
to reproduce the stability in peak position of the UV feature.
Having a coating to broaden the bump and eliminate correlations between
its width and its peak position can be considered a
second order effect, the overall shape of the grain being the first
order effect. Therefore, implicit in this model is the unlikely assumption
that {\it all} interstellar grains along every possible line of sight
have {\it exactly} the same shape. Any deviation in shape
shifts the peak position outside the observed range. Furthermore, variations
in shape introduce a correlation between the strength, the width, and the
peak position of the feature.
Since the width and peak position of the narrowest
interstellar UV features are narrower and shifted to smaller
wavenumbers, respectively, 
compared to those of Rayleigh graphite grains, Mathis had to
``tinker'' with the dielectric function $\epsilon_{\perp}$
of graphite tabulated by Draine (1985).
This is somewhat justified in view of the lack of
agreement between laboratory measurements carried out by various
authors in that spectral region (see e.g., Draine \& Lee 1984), but
it introduces additional free parameters. Note that virtually all of these
graphite optical constant measurements yield a plasmon resonance
for Rayleigh spheres at wavenumbers significantly larger than $4.60\mum1$.

Henrard et al.\ (1993) considered a model in which the bump carrier was
assumed to consist of small spherical onion shells of graphite. Again,
an unreasonable fine-tuning in the number of shells was required to
reproduce the peak position of the interstellar UV feature.

\section{Clustering effects on the graphite UV feature}\label{Graphite}

In this section we investigate whether clustering of grains composed
of graphite can lead to an increase of the width of the UV feature
without appreciably changing its peak position.

Draine (1988) and Draine \& Malhotra (1993) have confirmed the good
agreement between the so called ``1/3--2/3'' approximation
and DDA computations taking explicitly the anisotropy of graphite
into account. In this approximation extinction cross sections 
$C(\epsilon_{\parallel})$ and $C(\epsilon_{\perp})$ are obtained
separately for isotropic particles having the components of the
dielectric tensor parallel and perpendicular to the c-axis of
graphite, respectively. They are then combined
by taking 1/3 of the first contribution, and 2/3 of the second,
respectively. This procedure
is exact in the electrostatic limit for spheres, and is an excellent
approximation in the Rayleigh limit ($2\pi x a_{\rm eq} < 1$, where
$a_{\rm eq}$ is the radius of the equal-volume sphere). It is also
expected to be a good approximation for more complicated shapes in the
Rayleigh limit provided one assumes that the c-axis of graphite is
randomly oriented with respect to any symmetry axis that might be present
within the particles.

There is no consistent report of the interstellar UV feature being
polarized (i.e. the particles are not partially oriented elongated
particles). Therefore, in this paper, we assume that the particles
are randomly oriented.

It has been confirmed (Witt 1989; Witt et al.\ 1992; Calzetti et al.\ 1995)
that the extinction in the vicinity of the
interstellar UV feature is consistent with pure absorption 
(i.e., carrier with size in the Rayleigh limit).
This fact is exploited to simplify the orientational average by taking only
three mutually perpendicular orientations of the incident electric
field with respect to the particles. This
averaging procedure is also exact in the electrostatic limit and is 
a good approximation in the Rayleigh limit. An added advantage
of the Rayleigh limit is that one needs not worry about
size distributions of grains
since extinction is independent of size (except perhaps indirectly
through grains of different sizes possibly having different shapes).

It is of interest to see how the width, shape, and peak position
of the graphite UV feature change in the
case of ensembles of agglomerated spheres to simulate clustered or
irregularly shaped interstellar grains. Such a configuration can be handled by
a multiple Mie sphere code (Mackowski 1991; Rouleau 1996), provided
one assumes that the spheres, though touching, do no interpenetrate
each other, and are isotropic. 
Therefore, we have performed multiple Mie sphere computations to 
calculate extinction cross sections per unit volume, $C/V$, for 
various simple arrangements of spheres and plausible clustering 
models.

The two types of clusters considered are a cluster-cluster agglomerate
(CCA), and a compact cluster. A cluster-cluster
agglomerate is constructed from hierarchically joining together
clusters containing the same number of spheres, $N$,
in an $N=2,\;4,\;8,\;16,\dots$ progression
(see, e.g., Stognienko et al.\ 1995). These are rather fluffy and irregular
in shape (in the case $N=16$ considered here, the filling factor with respect
to a sphere enclosing the cluster is about 0.04). Another alternative
is to use a particle-cluster agglomerate (PCA) in which individual spheres
are added to the cluster with a sticking probability of unity. For
the small number of spheres used here, both CCA and PCA clusters
yield similar results and so PCA clusters are not included.
A compact cluster represents the tightest arrangement of spheres
possible, with each sphere touching many neighbours
(Rouleau \& Martin 1993; Rouleau 1996). Again, the spheres are not
allowed to interpenetrate. The filling factor of a compact cluster
is thus much larger than either CCA or PCA clusters (about 0.26 for
an $N=16$ cluster). This type of cluster could be the result of a low
sticking probability of the grains or a compaction of an initially looser
agglomerate after some disruptive event (like interstellar shocks).

The clusters are assumed to be in the Rayleigh limit.
The optical constants of Draine (1985) are used for the computations.
Qualitatively similar results are obtained for other choices of
optical constants of graphite.
A multipole expansion order up to 6 was necessary to obtain 
convergence for the $C/V$ results with the multiple Mie sphere 
calculations. 

\unitlength1cm
\begin{figure}[t]
\begin{picture}(0,6.2)\end{picture}
\includegraphics{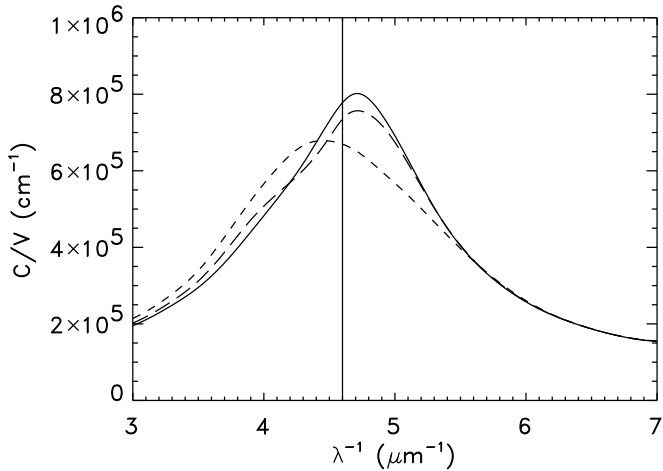}
\caption[]{\label{fig_3ways}
Cross section per unit volume, $C/V$, of randomly oriented 
$N=16$ CCA graphite clusters. 
Three different ways (see text) are used to treat the anisotropy of 
graphite: 
``1/3--2/3'' approximation (long dashed line), 
random assignment (solid line), 
and averaged dielectric function (short dashed line).
The vertical line marks the interstellar UV feature peak position
}
\end{figure}

There are three distinct ways to deal with the anisotropy of graphite 
in the multiple Mie sphere code.
The first way is to use the ``1/3--2/3'' rule, i.e.\ to average the 
extinction cross sections of the agglomerates with either 
$\epsilon_{\parallel}$ or $\epsilon_{\perp}$.
The second way is to randomly assign $\epsilon_{\parallel}$ or 
$\epsilon_{\perp}$ with probabilities 1/3 and 2/3 to the individual 
spheres forming the aggregates.
The third way is to average the components of the dielectric tensor 
and to use the resulting dielectric function in the multiple Mie 
sphere computations. 
These different approaches should be good approximations if we 
assume that there is no correlation between the orientations of 
the anisotropic spheres within the clusters and the direction of 
their individual c-axes.
This is true for the regions outside of the resonance.
However, it turns out that the results of the three methods differ 
from each other markedly in the bump region (Fig.~\ref{fig_3ways}).
E.g.\ for the CCA cluster the averaged dielectric function (third way) 
leads to a much broader feature ($\gamma = 2.20 \mum1$) at smaller 
wavenumbers ($x_{\rm max} = 4.48 \mum1$) compared to the results 
obtained with the ``1/3--2/3'' rule (results for other arrangements 
of spheres are discussed in detail below) 
and the random assignment (second way).
Furthermore, the feature obtained with the ``1/3--2/3'' rule shows a 
small shoulder in the long-wavelength wing.
The appearance of the shoulder is explained by the rather long 
linear chains of the strong $\epsilon_{\perp}$ ``oscillators''.
In case of the random assignment the chains are interspersed by the 
weak $\epsilon_{\parallel}$ ``oscillators'' which makes the feature 
less sensitive to the detailed cluster structure (see Sect.\ 
\ref{results_cca}).
The averaged dielectric function (third way) is in any case also more 
or less moderate compared to $\epsilon_{\perp}$ regardless of the used 
averaging procedure.
The feature position and width, however, certainly depend on the 
averaging procedure (the numbers given above are for the Bruggeman 
mixing rule; Bohren \& Huffman 1983).
Which way and which averaging procedure have to be used to get the 
best approximation of the $C/V$ of clusters of graphite spheres may 
be tested with extensive DDA calculations in the future, but this 
subject goes beyond the goals of this paper.
In any case, one should note that more detailed information about the 
arrangement of basic structured units (in respect to their individual 
c-axes) from laboratory data are required to get a more reliable basis 
for the calculations (see Rouzaud \& Oberlin 1989).
The remaining discussion in this section is restricted to the 
results obtained with the ``1/3--2/3'' rule for the various 
sphere configurations.

\begin{figure}[t]
\begin{picture}(0,6.2)\end{picture}
\includegraphics{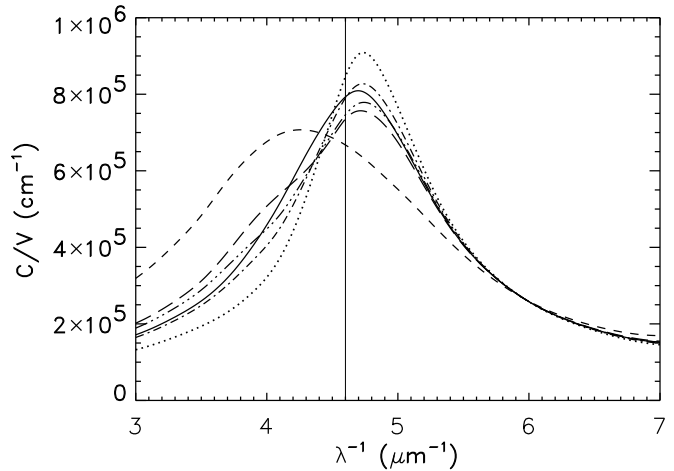}
\caption[]{\label{fig_graphite}
$C/V$ of 
a Rayleigh sphere (dotted line), 
a bisphere (dash-dotted line), 
a chain of three spheres (dash-triple-dot line),
an $N=16$ CCA cluster (long dashed line), 
an $N=16$ compact cluster (solid line), 
and CDE (short dashed line). 
All arrangements of spheres are randomly oriented, using the 
``1/3--2/3'' approximation to treat the anisotropy of graphite 
}
\end{figure}

Figure~\ref{fig_graphite} shows $C/V$ for a single sphere, a randomly
oriented bisphere, a randomly oriented linear chain of three touching
spheres, a CCA cluster containing 16 spheres, and
a compact cluster also containing 16 spheres.
Other random realizations of these clusters,
also containing 16 spheres, gave very similar results. In particular the
shoulder observed around $4 \mum1$ is reproduced in all CCA cluster
realizations. The appearance of the UV feature
looks very similar if only $C(\epsilon_{\perp})/V$ is plotted.
Note that in going from a single sphere to chains of increasing $N$, the
peak decreases in amplitude, while the extinction rises preferentially
in the long-wavelength wing of the feature. The feature of a CCA 
cluster is similar to that of a chain of three spheres, except that 
this extinction enhancement has developed into a shoulder. Also shown 
for comparison (short dashed line) is $C/V$ for a continuous 
distribution of ellipsoids (CDE; Bohren \& Huffman 1983), a widely 
used approximate clustering model.

\begin{table}[t]
\caption[]{\label{t-graphite}
Computed and fitted $C/V$ peak position (``PEAK'' and $x_{\rm max}$,
respectively) and width (``FWHM'' and $\gamma$, respectively)
for arrangements of graphite spheres (in $\mu {\rm  m}^{-1}$)}
\begin{tabular}{lrlrrr}
\noalign{\smallskip}\hline\noalign{\smallskip}
\multicolumn{1}{c}{cluster}
  & \multicolumn{1}{c}{PEAK} & \multicolumn{1}{c}{FWHM} 
  & \multicolumn{1}{c}{$x_{\rm max}$} & \multicolumn{1}{c}{$\gamma$}
  & \multicolumn{1}{c}{$\chi^2$}\\
\noalign{\smallskip}\hline\noalign{\smallskip}
1 sphere  & 4.743 & 1.21 & 4.762 & 1.06 &  67.6 \\
2 spheres & 4.735 & 1.48 & 4.738 & 1.31 &  74.6 \\
3 spheres & 4.742 & 1.73 & 4.731 & 1.52 & 225.2 \\
      CCA & 4.716 & 1.87 & 4.708 & 1.79 & 365.4 \\
  compact & 4.696 & 1.61 & 4.680 & 1.48 &  34.9 \\
      CDE & 4.268 & 2.46 & 4.332 & 2.40 &  21.4 \\
\noalign{\smallskip}\hline
\end{tabular}
\end{table}

Table~\ref{t-graphite} lists the peak position (``PEAK''),
full-width at half-maximum (``FWHM'') of the computed $C/V$,
as well as $x_{\rm max}$ and $\gamma$ derived from a fit similar in
form to Eq.~(\ref{is_fit}), with
fitting coefficients $c_1,\;c_2,\;c_3$ and $c_4$, using a slightly
modified version of the routine given in Fitzpatrick \& Massa (1990). 
To indicate the degree of quality of the fit, the $\chi^2$ value 
(in units of $10^6$) is also given, 
which is defined in the usual way with the weight for each $C/V$ 
data point (sampled at uniform intervals in wavenumber) set to the 
(arbitrary) value 1 except for the bump region $x=3.3$--$5.9\mum1$ 
where the weight is set to 2. 
Apart from the CDE case (which turns out to be a bad clustering model 
for the bump), the parameter $x_{\rm max}$ is quite stable
($\la 1.7\%$ relative variation), whereas $\gamma$ is substantially
increased (by $40-70\%$) in going from a single sphere to clusters.
These features satisfy qualitatively the observational constraints,
except for the fact that the peak position is wrong
($\sim 4.68$--$4.76\mum1$ compared to $4.60\pm0.04\mum1$). Note that there
is a significant difference between ``PEAK'' and
$x_{\rm max}$, and between ``FWHM`` and $\gamma$. This is due to the 
fact that a Drude profile fit is applied to
a profile that is not purely Drude-like, thus introducing a spurious
linear background (coefficients $c_1$, $c_2$) which
modifies both the peak position and the width of the feature.
The fit in the case
of the compact cluster is reasonable, which is indicated by the small
$\chi^2$ value in Table~\ref{t-graphite}, whereas the fit is somewhat
poorer in the case of the CCA cluster (due to
the presence of the shoulder), yielding a larger $\chi^2$.

The peak position is at a larger wavenumber than that observed for
the interstellar UV feature. This is why one has to resort to either shape
or size effects to shift the feature at the position of the interstellar
UV feature (see, e.g., Draine 1988; Mathis 1994).
Discounting the disagreement in peak position and concentrating
only on the qualitative aspect of the feature,
one can surmise that some irregularly shaped or clustered particles
{\it can} in fact broaden the feature without appreciably shifting
its peak position, as required for the purported interstellar UV
feature carrier. However, ``fluffy'' or elongated structures tend to
enhance the long wavelength wing of the feature, even introducing
additional structure in the feature (like the CCA cluster does).
Assuming this is a general feature of CCA clusters (see 
Sect.~\ref{Single_Lorentz}), it then appears that fluffy 
particles are 
severely constrained
as a possible ``topology'' of the interstellar UV feature carrier. 
A compact cluster
is an appealing alternative, since its overall spherical shape and
compactness suppresses the appearance of additional structure in the
feature, but still allows the feature to be broadened 
substantially compared to isolated
spheres. 
The broadening, however, appears to be insufficient to explain 
{\it all} the range of FWHM observed for the interstellar UV 
feature.
A further contributor to the broadening could be an intrinsic
variability in chemical
composition. This interpretation is consistent with the fact that
the narrowest interstellar UV features appear in all sorts of
environments, diffuse and dense alike, with an accompanying variation
in peak position over the whole observed range. We expect a clustering
of grains to occur primarily in denser regions. The ensuing broadening,
however, appears to contradict the observation that the broadest interstellar
UV features (observed in dense regions and thus presumably arising
from clustering)
are accompanied by a shift of $x_{\rm max}$ to {\it larger} values,
whereas here, it shifts to {\it smaller} values (see Table~\ref{t-graphite}).
But this does not necessarily need to be the case, as shown in the
following section.

\section{Single-Lorentz oscillator models}\label{Single_Lorentz}

Actually, in view of the widely assumed formation mechanisms
of the carbonaceous component responsible for
the UV bump, the highly anisotropic dielectric
function of planar graphite is unlikely to be a good model of the
purported dielectric function of the bump carrier. First, the exposure
of the carbonaceous interstellar grains to UV radiation, though conducive
to a ``graphitization'' of the material through dehydrogenation, is
unlikely to produce perfect graphite sheets as an end-product
(Sorrell 1990). There must
still be considerable defects in the structure, the topology of the
grains being a collection of randomly oriented graphitic crystallites in
a sp$^3$ bonding matrix (possibly containing some hydrogen).
This structure is more characteristic of amorphous carbon (especially 
UV processed or annealed hydrogenated amorphous carbon --- HAC; 
Fink et al.\ 1984; Mennella et al.\ 1995a,b).
Furthermore,
modification of the optical constants of graphite have already been
shown to be necessary in order to reproduce even the
most basic properties of the interstellar UV feature (Mathis 1994).
An intrinsic variability in chemical composition must also be
considered in order to satisfy the observational constraints.

Virtually all UV bumps observed to date
can be reproduced extremely well by a ``Drude'' profile.
This is the profile generated by a sphere whose optical properties are
characterized by a single-Lorentz oscillator model. Thus, in this section,
we consider a single-Lorentz oscillator model to approximate the
dielectric function hypothesized for the interstellar UV feature carrier,
and see how shape and clustering can affect the peak position, width,
and shape of the UV bump. We assume that variability in the
width is attributable to shape and clustering effects, as well as to 
intrinsic variations in chemical composition along different lines of sight.
Variability in the peak position, which is observed to be uncorrelated
with width, is attributed to variations in mean chemical composition alone.

The dielectric function
of a Lorentz oscillator is given by $\epsilon=1+x_p^2/[x_0^2-x^2-i\gamma_0 x]$,
where $x_p$, $x_0$, and $\gamma_0$, are the plasma frequency, the peak position,
and the damping constant, respectively, all in units of the inverse wavelength
($\mu {\rm  m}^{-1}$). A Drude model is characterized by $x_0=0$ and
usually describes metals (or semi-metals, like graphite). Through these
parameters,
the model can simulate the variability of the chemical composition of the
grains and provide some physical insight into the mechanisms involved.

Interstellar grains cannot be expected
to be perfectly smooth spheres in all environments. For example,
one may expect aggregation of the primary grains, especially in denser regions.
Furthermore, the grains could be characterized
by surface roughness and/or by porosity, as well as chemical inhomogeneities.
An interesting question is whether such shape and clustering effects
(neglecting chemical inhomogeneities within a given grain)
can conserve the Drude-like profile that is observed, along with the other
observational constraints.

\subsection{Combining a Lorentz oscillator model with shape and clustering}
The effect of shape and clustering on the scattering properties of
chemically homogeneous grains in the Rayleigh limit can be modelled
via a spectral
density, $g(L)$, of the geometric factor $L$, where $0 \le L \le 1$
(Bohren \& Huffman 1983; Fuchs 1987; Rouleau \& Martin 1991).
The requirements are that the zeroth and first moments of $g(L)$ are
unity and $1/3$, respectively. This approach is closely related
to the ``Bergman representation'' which is based on effective medium theory
in the case of a binary mixture (Stognienko et al.\ 1995). The two approaches
are equivalent if one assumes that one of the components is vacuum.
However, the spectral density $g(L)$ can be directly interpreted in terms
of shape and clustering of small grains, whereas the Bergman representation
deals with bulk material, and so gives only indirect information about the
scattering properties of small grains.

Combining a single-Lorentz oscillator model of dielectric function
$\epsilon=1+x_p^2/[x_0^2-x^2-i\gamma_0 x]$ with a model of shape
and clustering using $g(L)$, the extinction cross section per unit
volume can be written as
\begin{eqnarray}\label{CoV_gL}
\frac{C}{V}&=& 2\pi\,x\,{\rm Im}\int_0^1 \frac{g(L)\,dL}
{(\epsilon-1)^{-1}+L}\nonumber\\
&=&2\pi \int_0^1 \frac{x_p^2\gamma_0 g(L)\,dL}
{\left[ x-(x_0^2+x_p^2 L)/x \right]^2+\gamma_0^2}\nonumber\\
\end{eqnarray}
where ${\rm Im}$ denotes the imaginary part. The familiar form of
$C/V$ for spheres is obtained from $g(L)=\delta(L-1/3)$.
Note that $C/V$ using this model has the form
of the ``bump'' term in  Eq.~(\ref{is_fit}) if $g(L)=\delta(L-L_o)$.
In that case, the maximum $C/V$ occurs close to
\begin{equation} \label{xmax_Lo}
x_{\rm max}=[x_0^2+x_p^2 L_o]^{1/2}.
\end{equation}

One advantage of this type of representation is that a statistical
ensemble of grains of various shapes or clustering states can also be
represented by an average $g(L)$. Actually some mean $g(L)$ and
mean $\epsilon$ could
even describe qualitatively a statistical ensemble of grains with varying
shape {\it and} chemical composition, as long as there are no
correlations between the two.

The irregular grains considered here arise from the agglomeration of
spheres (i.e. the spheres can touch their neighbour at one point, but
they are not allowed to interpenetrate). In the case of agglomerates
of spheres, the spectral representation approach
brings about a considerable reduction in computer burden
compared to direct computations using a multiple Mie sphere code since
the problem has to be solved only once.
Results between the two approaches in the case of clusters of spheres
using the optical constants of graphite are very similar.

\begin{figure}[t]
\begin{picture}(0,6.3)\end{picture}
\includegraphics{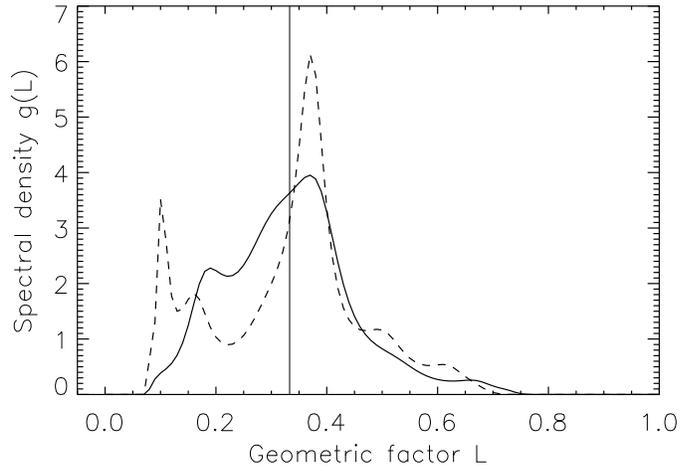}
\caption[]{\label{gL}
Spectral density for $N=16$ CCA clusters (dashed line) used in
Fig.~\ref{CoV_CCA}, and spectral density
for $N=16$ compact clusters (solid line) used in Fig.~\ref{CoV_cl}.
The vertical line represents the spectral density of spheres,
a delta function at $L=1/3$
}
\end{figure}

To compute the spectral density
$g(L)$ we use a code kindly provided by Hinsen \& Felderhof (1992).
This code computes ensemble-averaged electromagnetic interactions between 
identical spheres in the electrostatic limit.
As stated in Stognienko et al.\ (1995), the spectral density depends 
on the choice of the maximum order for the multipole expansion.
We present the spectral densities computed with multipole expansion 
order of 9 for the $N=16$ CCA clusters and the $N=16$ compact clusters 
in Fig.~\ref{gL}.
Note that a multipole expansion order of 6 is sufficient to obtain 
convergence for the $C/V$ results using Eq.~\ref{CoV_gL} with the 
optical constants of graphite. 

For our simple Lorentz model, we assign one of the smallest widths derived
from interstellar extinction curves, $\gamma=0.77$ (for HD~93028;
Fitzpatrick \& Massa 1986), to spheres. This might not be completely correct,
but the sphere is a useful reference shape. We assume here
that broader widths arise {\it exclusively} from shape and clustering
effects without any variation in chemical composition. We choose the peak
position as being the mean value, $x_{\rm max}=4.60\mum1$. This is an arbitrary
choice since the narrowest profiles ($\gamma<0.8$) span almost the
entire range of $x_{\rm max}$, from $4.58\mum1$ to $4.62\mum1$.
Other choices of $x_{\rm max}$ would yield qualitatively similar results,
but simply shifted in wavenumber. These would correspond to intrinsic
variations of the assumed chemical composition of the grains.

For spheres, a pure Drude profile is obtained. Due to shape and 
clustering effects, however, the synthetic bump computed from $g(L)$
may differ from a pure Drude profile. We wish to find out under which
conditions such clustering models can produce an increase in width
of the bump without appreciably changing its peak position or its
Drude-like shape.

\begin{figure}[t]
\begin{picture}(0,13.6)\end{picture}
\includegraphics{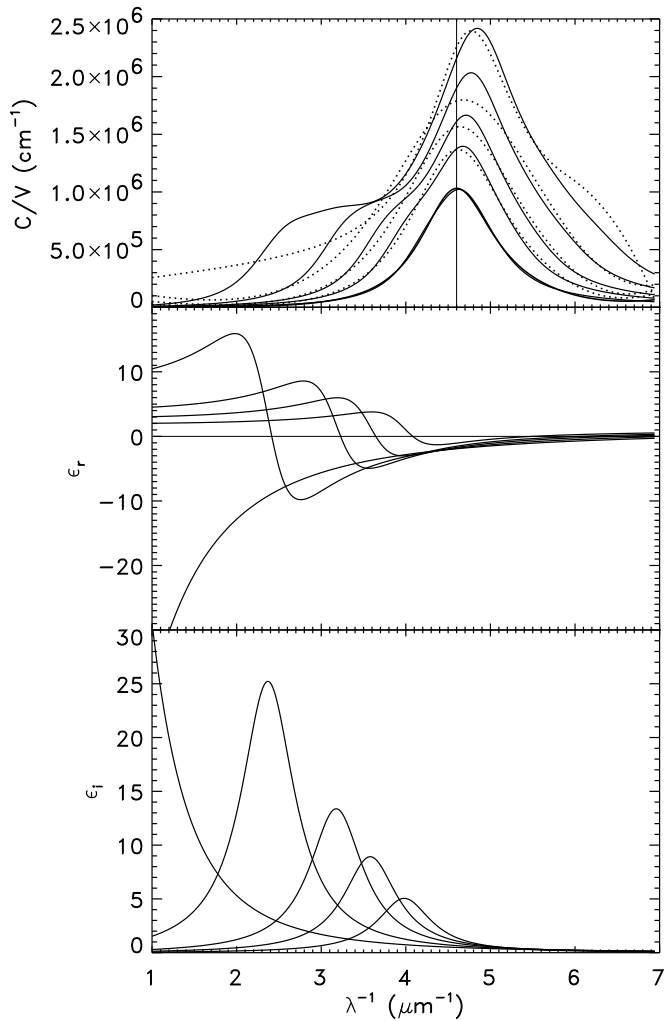}
\caption[]{\label{CoV_CCA}
$C/V$, $\epsilon_r$, and $\epsilon_i$, for various single-Lorentz 
oscillator models and the
$g(L)$ computed from $N=16$ CCA clusters. By order of increasing
amplitude, the models are given by $x_0=4.0,\;3.6,\;3.2,\;2.4,$
and 0.0 (solid lines). The dotted lines are the fit using
Eq.~(\ref{is_fit})
}
\end{figure}

\begin{table}[t]
\caption[]{\label{t-CCA}
Same as Table~\ref{t-graphite} but for single-Lorentz oscillator models
parameterized by $x_0$ in the case of $N=16$ CCA clusters
(PEAK~$=x_{\rm max}=4.60$ and FWHM~$=\gamma=0.77$ for spheres).
``MAX'' (cm$^{-1}$) is the computed peak value $(C/V)_{\rm max}$}
\begin{tabular}{rrrlrrr}
\noalign{\smallskip}\hline\noalign{\smallskip}
    \multicolumn{1}{c}{$x_0$}
  & \multicolumn{1}{c}{MAX} 
  & \multicolumn{1}{c}{PEAK}
  & \multicolumn{1}{c}{FWHM} 
  & \multicolumn{1}{c}{$x_{\rm max}$}
  & \multicolumn{1}{c}{$\gamma$} 
  & \multicolumn{1}{c}{$\chi^2$}\\
\noalign{\smallskip}\hline\noalign{\smallskip}
4.59 & 2.26E4 & 4.600 & 0.77 & 4.600 & 0.77 &   0.0 \\
4.50 & 2.21E5 & 4.601 & 0.78 & 4.600 & 0.78 &   0.0 \\
4.20 & 7.64E5 & 4.611 & 0.92 & 4.601 & 0.93 &  28.5 \\
4.00 & 1.02E6 & 4.629 & 1.06 & 4.605 & 1.09 & 139.8 \\
3.80 & 1.23E6 & 4.652 & 1.20 & 4.612 & 1.27 & 402.9 \\
3.60 & 1.40E6 & 4.674 & 1.33 & 4.624 & 1.47 &   --- \\
3.40 & 1.54E6 & 4.694 & 1.45 & 4.639 & 1.69 &   --- \\
3.20 & 1.67E6 & 4.712 & 1.55 & 4.657 & 1.91 &   --- \\
2.80 & 1.87E6 & 4.744 & 1.64 & 4.703 & 2.35 &   --- \\
2.40 & 2.03E6 & 4.771 & 1.60 & 4.744 & 2.65 &   --- \\
2.00 & 2.16E6 & 4.793 & 1.60 & 4.744 & 2.52 &   --- \\
1.60 & 2.26E6 & 4.811 & 1.62 & 4.723 & 1.78 &   --- \\
1.20 & 2.33E6 & 4.826 & 1.64 & 4.738 & 1.44 &   --- \\
0.80 & 2.38E6 & 4.836 & 1.65 & 4.752 & 1.31 &   --- \\
0.40 & 2.41E6 & 4.842 & 1.66 & 4.761 & 1.25 &   --- \\
0.00 & 2.42E6 & 4.844 & 1.66 & 4.764 & 1.24 &   --- \\
\noalign{\smallskip}\hline
\end{tabular}
\end{table}

\subsection{Results for CCA clusters}\label{results_cca}
Figure~\ref{CoV_CCA} shows
the extinction cross section per unit volume, $C/V$, and the real
and imaginary part of the dielectric function, $\epsilon_r$
and $\epsilon_i$, for various single-Lorentz oscillator
models using the $g(L)$ computed
from $N=16$ CCA clusters. The models were parameterized using $x_0$
and $x_p=[3(4.60^2-x_0^2)]^{1/2}$. This choice of $x_p$ gives rise to
a peak at $4.60\mum1$ for spheres ($L_o=1/3$ in Eq.~[\ref{xmax_Lo}]).
The damping constant is set to $\gamma_0=0.77\mum1$. For spheres,
this produces a Drude profile of width $\gamma=0.77\mum1$.

In order of increasing amplitude, results shown are for
$x_0=4.0,\;3.6,\;3.2,\;2.4,$ and 0.0 (solid lines).
Also shown for comparison are the fits using the procedure of 
Fitzpatrick \& Massa (1990; dotted lines). 
Note the considerable amount of structure in the profiles, except for 
the weakest model shown, $x_0=4.0$. 
The corresponding $g(L)$ is shown in Fig.~\ref{gL} (dashed line). 
A comparison between the two figures indicates that the $C/V$
of models with the smallest $x_0$ (``strong'' Lorentz models, i.e.
closest to a Drude model) are just
a convolution of $g(L)$ with a broadened Drude profile. Thus, for
``strong'' Lorentz models, extra peaks in the profile of $g(L)$ will
translate into structure in the $C/V$ curves. Conversely, for ``weak''
Lorentz models ($x_p^2 L \ll x_0^2$ in Eq.~[\ref{CoV_gL}])
the profile will be close to a ``Drude'' profile, irrespective of the
form of $g(L)$.

Table~\ref{t-CCA} is similar to Table~\ref{t-graphite}, but for the
one-Lorentz oscillator models using the $g(L)$ of CCA clusters.
In the present case, $x_{\rm max}$ and $\gamma$ are meaningless for
$x_0 \la 3.8$ since the curves are poorly represented by a ``Drude''
profile (thus being inconsistent with observations of actual interstellar
extinction curves). Again, note the significant discrepancies between 
``PEAK'' and $x_{\rm max}$, and between ``FWHM'' and $\gamma$ because 
of the spurious fitted linear rise. Only $\chi^2$
values of less than 500 are listed, corresponding to models that are
acceptable. Note that, though the curves are not strictly Drude-like,
``PEAK'' shifts to {\it larger} values
compared to that of spheres (4.60$\mum1$) for broader profiles,
precisely what is observed in denser interstellar environments. This is
related to the fact that the $g(L)$ profile peaks at a value of $L$
larger than 1/3 (the value of $L$ for spheres, indicated by a vertical
line in Fig.~\ref{gL}).

The ratio of carbon locked up in interstellar grains relative
to hydrogen, $N_{\rm C}/N_{\rm H}$, is thought be be around $\sim 100$
part-per-million (ppm)
by number (Cardelli et al.\ 1996). Assuming the bump grains have a density of
2~g~cm$^{-3}$, then only the ``weakest'' Lorentz models 
($x_0 \ga 4.5$) require more carbon than available 
(cf.\ the values of $(C/V)_{\rm max})$ in Table~\ref{t-CCA}).
As expected, putting all the carbon in a single extinction feature 
requires only modest amounts of the element.

Thus cluster-cluster agglomeration (i.e. fluffy interstellar grains)
{\it can} reproduce the observational constraints relating to the
interstellar UV feature, provided the bump grains are described by a
single-Lorentz oscillator model with $x_0 \sim 3.8$--$4.5\mum1$. 

\begin{figure}[t]
\begin{picture}(0,13.6)\end{picture}
\includegraphics{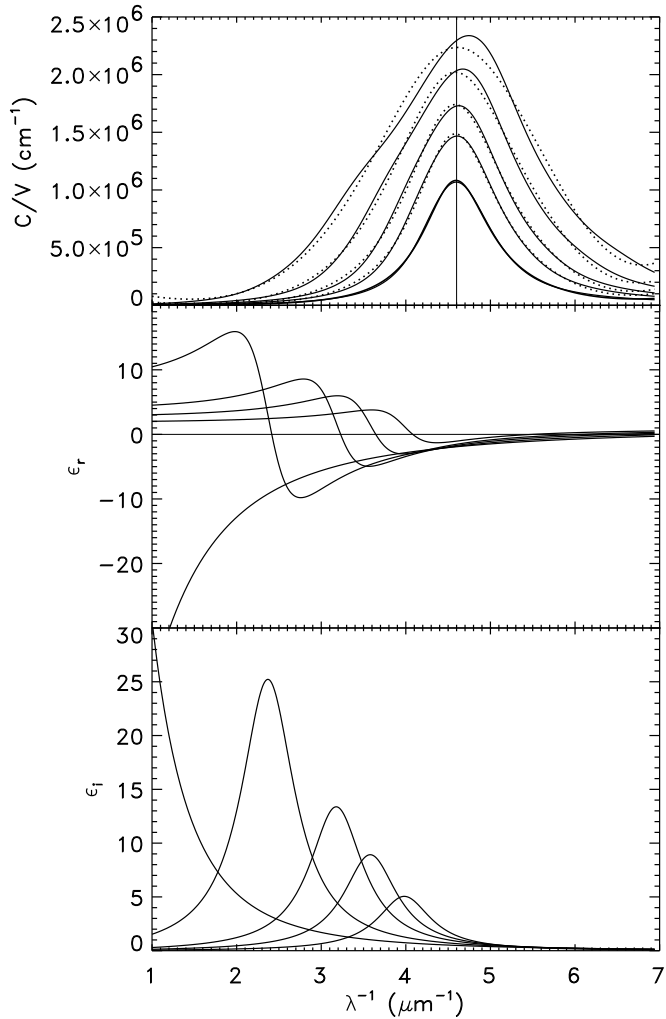}
\caption[]{\label{CoV_cl}
Same as Fig.~\ref{CoV_CCA}, but for a $g(L)$ computed from
$N=16$ compact clusters (solid line in Fig.~\ref{gL})
}
\end{figure}

\begin{table}[ht]
\caption[]{\label{t-cl}
Same as Table~\ref{t-CCA}, but in the case of $N=16$ compact clusters}
\begin{tabular}{rrrlrrr}
\noalign{\smallskip}\hline\noalign{\smallskip}
    \multicolumn{1}{c}{$x_0$}
  & \multicolumn{1}{c}{MAX} 
  & \multicolumn{1}{c}{PEAK}
  & \multicolumn{1}{c}{FWHM} 
  & \multicolumn{1}{c}{$x_{\rm max}$}
  & \multicolumn{1}{c}{$\gamma$} 
  & \multicolumn{1}{c}{$\chi^2$}\\
\noalign{\smallskip}\hline\noalign{\smallskip}
4.59 & 2.25E4 & 4.600 & 0.77 & 4.600 & 0.77 &   0.0 \\
4.50 & 2.23E5 & 4.600 & 0.78 & 4.600 & 0.78 &   0.0 \\
4.20 & 7.86E5 & 4.597 & 0.88 & 4.598 & 0.88 &  12.0 \\
4.00 & 1.07E6 & 4.596 & 0.99 & 4.595 & 0.99 &  54.6 \\
3.80 & 1.29E6 & 4.600 & 1.10 & 4.592 & 1.12 & 125.7 \\
3.60 & 1.47E6 & 4.607 & 1.20 & 4.590 & 1.25 & 213.9 \\
3.40 & 1.61E6 & 4.616 & 1.31 & 4.590 & 1.39 & 318.1 \\
3.20 & 1.73E6 & 4.626 & 1.41 & 4.591 & 1.53 & 448.2 \\
2.80 & 1.91E6 & 4.650 & 1.60 & 4.597 & 1.83 &   --- \\ 
2.40 & 2.05E6 & 4.673 & 1.76 & 4.609 & 2.12 &   --- \\
2.00 & 2.15E6 & 4.693 & 1.89 & 4.627 & 2.40 &   --- \\
1.60 & 2.22E6 & 4.712 & 1.99 & 4.649 & 2.67 &   --- \\
1.20 & 2.27E6 & 4.726 & 2.07 & 4.671 & 2.88 &   --- \\
0.80 & 2.31E6 & 4.736 & 2.13 & 4.691 & 3.05 &   --- \\
0.40 & 2.33E6 & 4.742 & 2.16 & 4.704 & 3.15 &   --- \\
0.00 & 2.34E6 & 4.744 & 2.17 & 4.709 & 3.19 &   --- \\
\noalign{\smallskip}\hline
\end{tabular}
\end{table}

\subsection{Results for compact clusters}
We also computed $C/V$ using Eq.~(\ref{CoV_gL}) for a $g(L)$ corresponding
to an ensemble of compact clusters containing $N=16$ spheres.
Figure~\ref{CoV_cl} is similar to Fig.~\ref{CoV_CCA}, but for
an ensemble of $N=16$ compact cluster models. The corresponding $g(L)$
is shown in Fig.~\ref{gL} (solid line). Note that the $g(L)$ 
of compact clusters also
peaks at $L>1/3$, but that it is more ``compact'' and contains less structure
than the $g(L)$ of CCA clusters. This translates into $C/V$ profiles
that are closer to a Drude profile, except again for ``stronger'' Lorentz
models ($x_0 \la 3.6$). This emphasizes how tightly constrained
the shape and clustering of grains described by
the stronger single-Lorentz oscillator models are in terms of their allowable
$g(L)$ profile. Table~\ref{t-cl} lists the same parameters as
Table~\ref{t-CCA} for the various compact clusters using these 
Lorentz models. Models with $3.2 \la x_0 \la 4.5$ satisfy most
of the observational constraints. Only models with $x_0 \ga 4.5$
violate the cosmic carbon abundance constraint. Note again the trend
of larger FWHM's usually leading to larger values of ``PEAK'', as required
observationally.

\section{Two-Lorentz oscillator models}\label{Two_Lorentz}
The simple approach taken in the previous section -- to model
the optical properties of the interstellar UV feature carrier
with a single Lorentz oscillator -- may appear somewhat unrealistic.
It assumes that the bump carrier does not contribute at all to
the FUV rise. Actually, along some lines of sight (many passing
through Orion), the FUV rise is observed to be very small or
virtually absent, even though the interstellar UV feature is
still present, albeit weaker than usual. But all carbonaceous
materials (including diamond) are expected to 
give rise to a peak at roughly 10--15$\mum1$ due to the
$\sigma$--$\sigma^*$ electronic transition (Mennella et al.\ 1995a;
Fink et al.\ 1984).
Moreover, any plausible material in the small particle limit will 
produce strong absorption at $x > 6\mum1$.

To our knowledge, no physical mechanism can suppress the 
$\sigma$--$\sigma^*$ transition while keeping intact the 
$\pi$--$\pi^*$ transition responsible for the UV feature. 
Thus, the assumption that the UV bump carrier is carbonaceous in 
nature appears 
at first glance to be difficult to reconcile with the very small FUV 
rise observed in a number of cases.
In this section, we ignore this difficulty and assume that the
dielectric function of the interstellar UV feature carrier is the
sum of two Lorentz oscillators -- one producing the UV bump, and the
other producing the FUV curvature-- to see how the conclusions of the
preceding section are modified.

One important distinction must be made here. If two grain populations
are unrelated, then one can add their cross sections, like is
implicit in the decomposition of Eq.(\ref{is_fit}). For example,
adding an extra linear background or an extra FUV curvature will affect
only their corresponding fitting parameters in Eq.~(\ref{is_fit}), but
will leave the other fitting parameters unchanged. However, if the
UV bump and the FUV curvature arise from the same parent material,
then their respective contributions to the total {\it dielectric function}
must be added instead, and {\it then} only can the cross section be
computed assuming a certain grain shape. These various contributions to the
dielectric function do not necessarily add up linearly when translated
into cross sections, unless $\epsilon_i$ is small. Therefore,
adding an extra Lorentz oscillator at larger $x$ (in $\epsilon$ space)
to an initial one that produces a UV bump well characterized by a
``Drude'' profile (for example, assuming a spherical shape)
might actually result into a bump that is no longer Drude-like.
As a result of this process, all fitting parameters may be affected
in a non-trivial way.

To investigate these effects we approximate such a
dielectric function as the sum of two Lorentz oscillators, giving rise to
features at about $4.6\mum1$ (the UV bump) and 10--15$\mum1$ (the FUV bump),
respectively. This guarantees that the total dielectric function of the
material satisfies Kramer-Kronig relations (Bohren \& Huffman 1983).
Of course, more complicated dielectric functions are also possible, but
then the number of free parameters becomes rapidly excessive, and the
physical interpretation becomes less straightforward.

\begin{figure}[t]
\begin{picture}(0,13.6)\end{picture}
\includegraphics{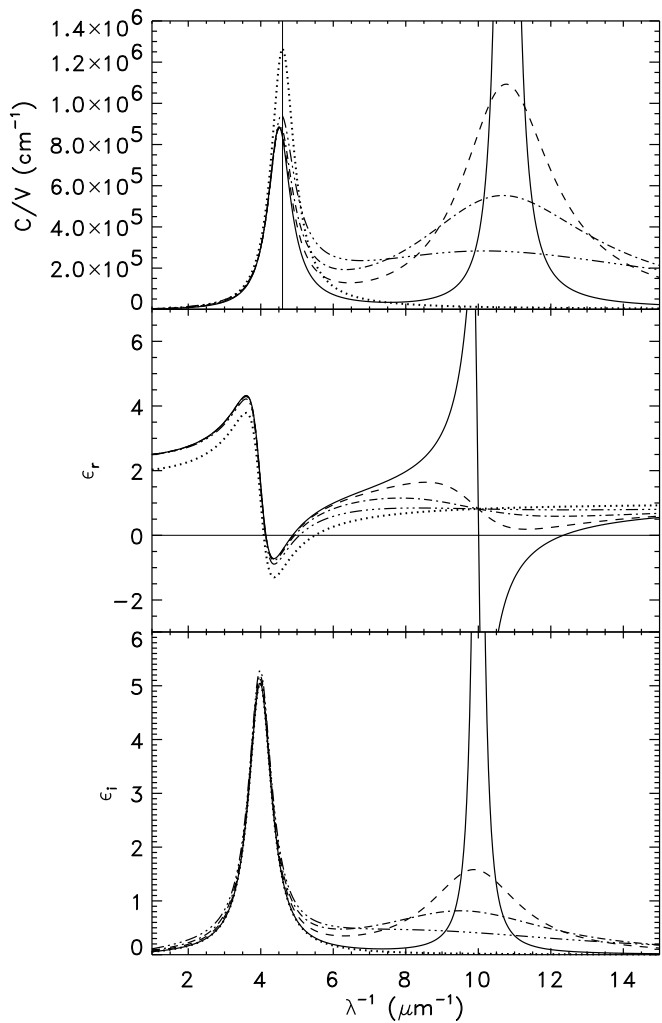}
\caption[]{\label{two_Lorentz_width}
Same format as Figs.~\ref{CoV_CCA} and \ref{CoV_cl}, but for various
Lorentz oscillator models 
and for spheres. 
Results shown are for a single-UV-Lorentz oscillator model with 
$x_{01}=4.0$, $x_{p1}=3.934$, $\gamma_{01}=0.77$ (dotted lines), 
and for models involving an extra FUV Lorentz oscillator with 
$x_{02}=10.0$, $x_{p2}=6.833$, and $\gamma_{02}=
0.3$ 
(solid lines),
3.0 (dashed lines), 6.0 (dash-dotted lines), and 12.0 (dash-triple-dot lines)
}
\end{figure}

The effects of this extra Lorentz oscillator on the UV bump are illustrated
in Fig.~\ref{two_Lorentz_width}. The UV oscillator is given by the
parameters $x_{01}=4.0$, $x_{p1}=3.934$, and $\gamma_{01}=0.77$, which on its
own yields a 
$C/V$ 
peak at 4.60$\mum1$ assuming a sphere (dotted line).
The FUV oscillator is given by $x_{02}=10.0$, $x_{p2}=6.833$, and various
widths, which on its own yields a 
$C/V$ 
peak at $10.75\mum1$ assuming a sphere.
This could be viewed as a very crude model of the $\sigma$--$\sigma^*$ peak
observed for HAC in laboratory extinction experiments
(Mennella et al.\ 1995a). The solid, dashed, dash-dotted, and dash-triple-dot
lines are for $\gamma_{02}=0.3,\;3.0,\;6.0$, and 12.0, respectively
(also assuming spheres). 
The main effect of the FUV oscillator is to decrease the strength 
of the UV bump, more or less independently of the width of the 
FUV bump (keeping the FUV oscillator strength constant).
Another effect is to shift the peak position of the UV bump to
smaller wavenumbers ($4.506,\;4.519,\;4.536$, and 4.573$\mum1$, for
$\gamma_{02}=0.3,\;3.0,\;6.0$, and 12.0, respectively). An extra broadening
also occurs, but only on the larger $x$ wing of the bump.
Changing the width of the FUV oscillator without changing its strength
affects the FUV curvature dramatically. Actually, this suggests an
explanation for the quasi-absence of FUV curvature observed along some lines
of sight: if the FUV bump is extremely narrow (or shifted to large
enough wavenumbers), then the FUV rise may be virtually absent (solid line).
This is also obtained for very wide FUV bumps, but then
there is a ``jump'' in the linear component of the extinction across
the UV bump (dash-triple-dot line).
However, Fitzpatrick \& Massa (1988) have shown
that the linear background of interstellar extinction curves, once the
fitted bump has been subtracted out, is very smooth across the bump,
with no evidence of jumps or changes of slope. Therefore extremely
broad FUV bumps are more or less ruled out as an explanation of the
lack of a FUV curvature concurrent with the presence of a UV bump.

The main effect of adding such a background to the dielectric function
is to increase $\epsilon_r$ and, to a lesser extent, $\epsilon_i$,
shifting the plasmon condition ($\epsilon_r < 0$) to smaller wavenumbers.
Note how little the dielectric function
of the UV oscillator is changed for smaller wavenumbers when changing
the width of the FUV oscillator (keeping the strength constant).

When a shape different from a sphere is assumed (either a CCA or a compact
cluster), as a result of this extra background, the peak position is
shifted to {\it smaller} wavenumbers, contrary to what was obtained in
the previous section. A moderate increase in FWHM is also obtained.
These results are similar to what was obtained in the case of
graphite (Table~\ref{t-graphite}).

\begin{figure}[t]
\begin{picture}(0,6.3)\end{picture}
\includegraphics{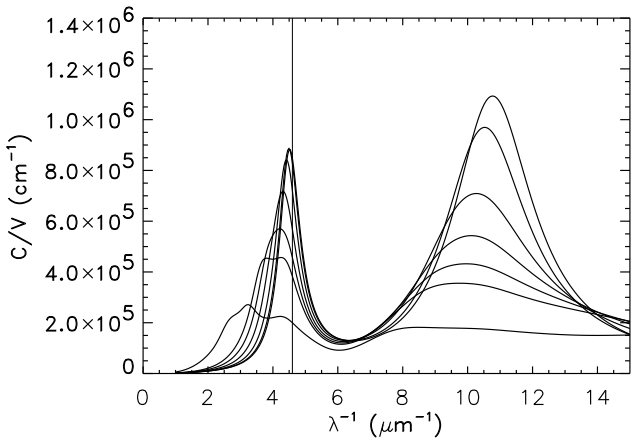}
\caption[]{\label{two_Lorentz_Mie}
$C/V$ of Mie spheres for various radii $a$ using the dielectric function
with $\gamma_{02}=3.0$ in Fig.~\ref{two_Lorentz_width} (dashed lines). 
In order of decreasing strength, $C/V$ for radius
$a$ in the Rayleigh limit, and $a=0.01,\;0.02,\;0.03,\;0.04,\;0.05,$
and 0.1~$\mu$m, respectively
}
\end{figure}
\subsection{The effect of scattering on the FUV curvature}
Another mechanism that can flatten the FUV bump is scattering by
larger grains. But scattering is likely to affect the UV feature as
well (mainly shifting it to smaller wavenumbers -- which is not observed
in all interstellar cases), unless a size distribution could be such that
it would affect only wavenumbers larger than $\sim 4.6\mum1$. However, this
is unlikely, as shown in Fig.~\ref{two_Lorentz_Mie}. Plotted (in order
of decreasing strength) are the $C/V$ Mie curves for spheres with radius
$a$ in the Rayleigh limit, and $a=0.01,\;0.02,\;0.03,\;0.04,\;0.05,$
and 0.1~$\mu$m, respectively, using the dielectric function
with $\gamma_{02}=3.0$ in Fig.~\ref{two_Lorentz_width} (dashed lines). 
Note that the UV bump is strongly affected
in peak position and shape for $a \ga 0.03~\mu$m, even though the
FUV peak is still quite large (with much FUV curvature).
The size which produces a FUV bump that is
essentially flat, $a=0.1~\mu$m (lowest curve), also yields a UV bump
that is virtually unrecognizable. Here, the UV peak is more strongly
affected than the FUV peak because the dielectric function in the
UV range is closer to a plasmon condition (see
Fig.~\ref{two_Lorentz_width}). Therefore, a size distribution of
grains (weighting these curves by the volume of individual grains) is
unlikely to flatten the FUV rise without severely affecting the shape
and peak position of the UV bump as well.

\subsection{Synthetic interstellar extinction curves}
\begin{figure}[t]
\begin{picture}(0,13.6)\end{picture}
\includegraphics{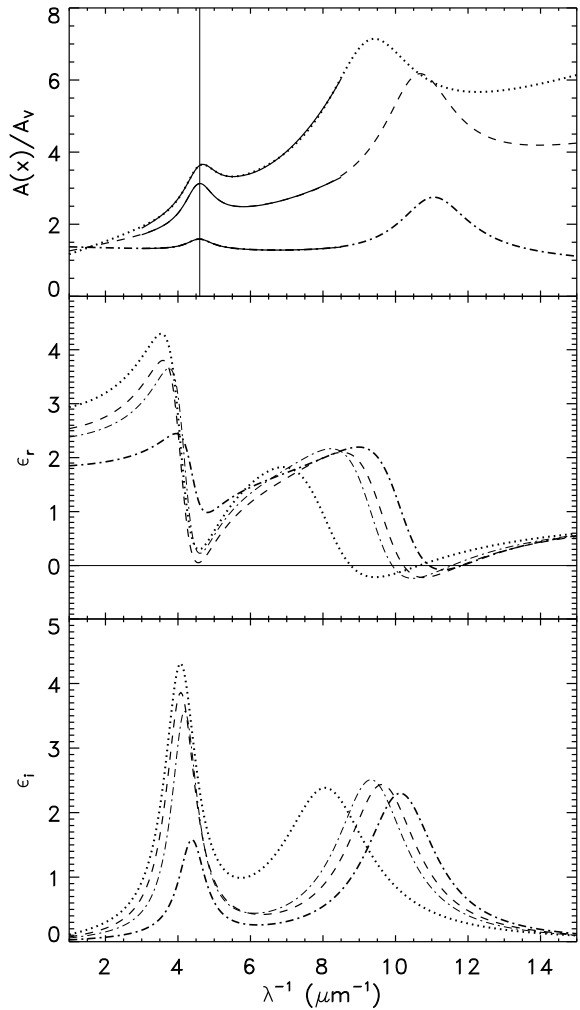}
\caption[]{\label{aoav_isc}
$A(x)/A_{\rm V}$, $\epsilon_r$, and $\epsilon_i$, for various Lorentz 
oscillator models assuming spheres. The parameters of the models, listed
in Table~\ref{t-Lorentz-isc}, were chosen to fit actual interstellar
curves (solid lines). Models are for HD~204827 (dotted lines), HD~229196
(dashed lines), and HD~37023 ($\theta^1$~Ori~D; dash-dotted lines: 
thick---$N_{\rm C}/N_{\rm H} = 90$~ppm, thin---45~ppm)
}
\end{figure}

A direct comparison between the fitting parameters of interstellar
curves and some synthetic ones can be useful to assess
the qualities of the Drude profile for the UV bump and the 
term $f_{\rm FUV}(x)$ for the FUV curvature when two-Lorentz 
oscillator models are considered. To make such a comparison, a few
simplifying assumptions are required. The total optical depth is
assumed to arise from two contributions, one from grains producing most
of the linear rise between 3 and 8.5$\mum1$ (labelled $l$) and one from
the grains producing the bump, the rest of the linear rise,
and the FUV curvature (labelled $b$). The component $l$ is assigned to
silicate grains. The lack of correlation between the
linear rise parameters and the bump parameters suggests that the
grains responsible for the UV bump do not contribute appreciably to the
linear rise (Jenniskens \& Greenberg 1993). Thus assigning most of the
linear rise to silicate grains is reasonable. The different slopes
observed in the interstellar extinction curves (see, e.g., Fig.~\ref{aoav_isc})
could be assumed to arise from
different size distributions of silicate grains (with larger mean grain
sizes producing flatter curves). For the sake of simplicity,
we shall not be concerned here with the detailed modelling of this component,
but will simply assume that its $C/V$ can be described 
as a simple linear rise in the range 3--8.5$\mum1$. For the carbonaceous
component $b$, a key normalization parameter is the carbon to hydrogen ratio
(by number), $N_{\rm C}/N_{\rm H}$.

Assuming $A_{\rm V} = 1.7 \times 10^{-22} N_{\rm H} R_{\rm V}$ (Mathis 1994) and
using $A(x)=1.086 (N_l C_l+N_b C_b)$, where $N_j$ is the column density
of dust and $C_j$ is the cross section of component $j$ (where $j=l,\,b$),
we have from the definition of $k(x)$,
\begin{eqnarray} \label{def_k}
k(x) &=& R_{\rm V} \frac{A(x)}{A_{\rm V}}-R_{\rm V}\nonumber\\
&=& \frac{1.086}{1.7 \times 10^{-22}}\frac{N_b}{N_{\rm H}}
\left[\frac{N_l}{N_b}C_l(x)+C_b(x)\right]-R_{\rm V}, \nonumber\\
\end{eqnarray}
where $R_{\rm V}=A_{\rm V}/E$(B-V) is the total-to-selective extinction ratio in
the V band. We have $N_b \simeq 12 m_{\rm H} N_{\rm C}/(\rho_b V_b)$, where $m_{\rm H}$
is the mass of the hydrogen atom, and $\rho_b$ and $V_b$ 
are the density and the mean volume of the grains, respectively.
The density of bump grains is again assumed to be $2\,{\rm g}{\rm cm}^{-3}$.
The computed cross section per unit volume of the bump grains using
two Lorentz oscillators (i.e. seven free parameters: $x_{0j}$,
$x_{pj}$, $\gamma_{0j}$, $j=1,2$, and $N_{\rm C}/N_{\rm H}$) is fitted by
\begin{equation} \label{CoV_fit}
    \frac{C_b(x)}{V_b} = c_{b1}+c_{b2} x
+\frac{c_{b3}}{(x-x_{\rm max}^2/x)^2+\gamma^2}
+c_{b4} f_{\rm FUV}(x)\nonumber\\
\end{equation}
between 3 and 8.5$\mum1$.
Assuming that $C_l(x)$ is linear with respect to $x$,
Eq.~(\ref{def_k}) can be rewritten as
\begin{equation}\label{synth_k}
k(x) = R_{\rm V} C_0 \left[c_{l1}+c_{l2} x+\frac{C_b(x)}{V_b}\right]-R_{\rm V},
\end{equation}
where $c_{l1}=(a_1+R_{\rm V})/(R_{\rm V} C_0)-c_{b1}$, 
$c_{l2}=a_2/(R_{\rm V} C_0)-c_{b2}$,
and $R_{\rm V} C_0 \simeq 0.064 N_{\rm C}/N_{\rm H}$.

Interstellar extinction curves along specific lines of sight were modelled
using this two-Lorentz oscillator model assuming Rayleigh spheres over
the whole range. Using the same optical constants and the same normalization,
results for CCA clusters and compact clusters were also computed to
see how the bump parameters changed. Uncertainties related to the
extrapolation beyond the observed range (3--8.5$\mum1$) do not warrant
taking the extra complication of scattering into account.
The value of $N_{\rm C}/N_{\rm H}$ was set more or less arbitrarily
to 90~ppm (i.e. slightly below the current derived limit) 
and to 45 for the line of sight where a weak UV bump is observed (case 
of HD~37023). 
Smaller values than 90~ppm for the other two lines of sight 
would require more extreme Lorentz oscillator parameters and extinction
curves may exhibit some substructure (see Fig.~\ref{CoV_CCA}), and so
it might be more difficult to obtain a good fit.

Figure~\ref{aoav_isc} shows $A(x)/A_{\rm V}$ for spheres
using $\epsilon_r$ and $\epsilon_i$ (shown in the middle
and bottom panels, respectively). The dielectric function was computed
from the parameters of two-Lorentz oscillator models. These parameters,
listed in Table~\ref{t-Lorentz-isc}, were chosen to reproduce (assuming
spheres) the extinction along three lines of sight spanning
a wide range in FUV curvature (Fitzpatrick \& Massa 1988; 1990).
The lines of sight are in the direction of HD~204827 (top solid line, modelled
by the dotted lines), HD~229196 (middle solid line, modelled by the dashed
lines), and HD~37023 ($\theta^1$~Ori~D; lower solid line, modelled by the
dash-dotted lines). 
The values of $R_{\rm V}$ were taken from Cardelli et al.\ (1989). 
In Table~\ref{t-isc}, the resulting fitting
parameters obtained for spheres, CCA clusters and compact clusters
are compared to those
of these interstellar extinction curves. 
As can be shown by the $\chi^2$ values
(relative to $C/V$, not $A/A_{\rm V}$; in units of $10^6$), the fit is excellent. 
Surprisingly, the
fit can actually {\it improve} in going from spheres to clusters
(case of HD~204827). This is mainly due to a better fit of the FUV
curvature, not the UV bump. 
In case of HD~37023 the fit becomes slightly worse (due to a slight 
FUV curvature mismatch) in reducing $N_{\rm C}/N_{\rm H}$ from 90 
to 45~ppm. 
Both oscillators are shifted to smaller wavenumbers in case of 
$N_{\rm C}/N_{\rm H} = 45$~ppm, and, as expected, the fitted UV 
oscillator appears to be much stronger (larger $x_{p1}$) than 
for the case of 90~ppm.
Interestingly, $\gamma_{01}$ is even more reduced in case of 45~ppm.
At any rate, the $\chi^2$ values are usually
smaller than in the single-Lorentz oscillator case (see entries $x_0=4.0$ in
Tables~\ref{t-CCA} and \ref{t-cl}). Thus it appears that an extra background
under the UV bump (due to the FUV oscillator contribution to the UV
oscillator) makes the bump profile less sensitive to shape and clustering.

\begin{table*}
\caption[]{\label{t-Lorentz-isc} Parameters of two-Lorentz oscillator models 
to reproduce interstellar extinction curves along selected lines of sight}
\begin{tabular}{lrrrrrrrrcr}
\noalign{\smallskip}\hline\noalign{\smallskip}
  & \multicolumn{1}{c}{$x_{01}$}
  & \multicolumn{1}{c}{$x_{p1}$}
  & \multicolumn{1}{c}{$\gamma_{01}$}
  & \multicolumn{1}{c}{$x_{02}$}
  & \multicolumn{1}{c}{$x_{p2}$}
  & \multicolumn{1}{c}{$\gamma_{02}$}
  & \multicolumn{1}{c}{$c_{l1}$$\times$$10^6$} 
  & \multicolumn{1}{c}{$c_{l2}$$\times$$10^6$} 
  & \multicolumn{1}{c}{$N_{\rm C}/N_{\rm H}$ (ppm)}
  & \multicolumn{1}{c}{$R_{\rm V}$}\\
\noalign{\smallskip}\hline\noalign{\smallskip}
HD~204827&4.100&4.200&1.060& 8.200&7.400&2.900&0.369&   0.151&90&2.60\\
HD~229196&4.105&3.890&0.980& 9.700&7.400&2.350&0.570&   0.105&90&3.12\\
HD~37023 &4.407&2.425&0.890&10.200&7.400&2.350&1.265&$-$0.028&90&5.23\\
\noalign{\smallskip}
HD~37023 &4.203&3.553&0.868& 9.400&7.400&2.350&2.522&$-$0.052&45&5.23\\
\noalign{\smallskip}\hline
\end{tabular}
\end{table*}

\begin{table*}
\caption[]{\label{t-isc} Fitting parameters of synthetic interstellar 
extinction curves compared to actual ones}
\begin{tabular}{lcrrrrrrr}
\noalign{\smallskip}\hline\noalign{\smallskip}
  & \multicolumn{1}{c}{$N_{\rm C}/N_{\rm H}$ (ppm)}
  & \multicolumn{1}{c}{$x_{\rm max}$}
  & \multicolumn{1}{c}{$\gamma$}
  & \multicolumn{1}{c}{$a_1$}
  & \multicolumn{1}{c}{$a_2$}
  & \multicolumn{1}{c}{$a_3$}
  & \multicolumn{1}{c}{$a_4$}
  & \multicolumn{1}{c}{$\chi^2$}\\
\noalign{\smallskip}\hline\noalign{\smallskip}
HD~204827&   & 4.623 & 1.070 &$-$1.521 &   1.219 & 3.201 & 0.899 &    \\
sphere   & 90& 4.623 & 1.075 &$-$1.521 &   1.219 & 3.200 & 0.882 &46.9\\
CCA      & 90& 4.553 & 1.166 &$-$1.575 &   1.235 & 3.701 & 0.871 &15.1\\
compact  & 90& 4.575 & 1.127 &$-$1.561 &   1.231 & 3.495 & 0.880 &15.5\\
\noalign{\smallskip}\hline\noalign{\smallskip}
HD~229196&   & 4.581 & 0.990 &$-$0.179 &   0.728 & 3.407 & 0.233 &    \\
sphere   & 90& 4.582 & 0.991 &$-$0.179 &   0.728 & 3.410 & 0.234 & 9.2\\
CCA      & 90& 4.541 & 1.128 &$-$0.213 &   0.723 & 4.263 & 0.271 &14.6\\
compact  & 90& 4.552 & 1.078 &$-$0.211 &   0.726 & 3.951 & 0.257 &10.0\\
\noalign{\smallskip}\hline\noalign{\smallskip}
HD~37023 &   & 4.594 & 0.878 &   1.883 &$-$0.083 & 1.215 & 0.153 &    \\
sphere   & 90& 4.594 & 0.879 &   1.883 &$-$0.083 & 1.215 & 0.153 & 2.1\\
CCA      & 90& 4.579 & 0.903 &   1.879 &$-$0.083 & 1.289 & 0.164 & 2.5\\
compact  & 90& 4.583 & 0.895 &   1.880 &$-$0.083 & 1.263 & 0.160 & 2.4\\
\noalign{\smallskip}
sphere   & 45& 4.594 & 0.878 &   1.883 &$-$0.083 & 1.215 & 0.153 &18.1\\
CCA      & 45& 4.558 & 0.977 &   1.870 &$-$0.084 & 1.465 & 0.173 &20.8\\
compact  & 45& 4.568 & 0.941 &   1.872 &$-$0.083 & 1.373 & 0.166 &18.9\\
\noalign{\smallskip}\hline
\end{tabular}
\end{table*}

Unfortunately, given the already large number of free parameters (7),
the models are not unique. But they give a qualitative idea of what
the optical constants of {\it actual} bump grains
might look like in various environments. Note that for curves exhibiting less
FUV curvature, the second Lorentz oscillator must be shifted to larger
$x$, otherwise the fit can be very poor in the FUV portion of the observed
range. The FUV curvature observed in the direction
of HD~204827 is so large ($a_4=0.899$) that no FUV Lorentz oscillator
with $x_{02} \simeq 10\mum1$ can reproduce it (these tend to give
$a_4 \la 2.2$). These large curvatures can only be generated by bringing
the FUV oscillator closer to the UV oscillator ($x_{02} \sim 8\mum1$). The fit
is slightly poorer in that case for spheres since the tail of the FUV
Lorentz oscillator significantly differs from the shape of this curvature.
The bump peak position, $x_{\rm max}$, is sensitive mainly to $x_{01}$
(a $1\%$ variation in $x_{01}$ produces a $0.8\%$ variation in $x_{\rm max}$
-- which is a substantial part of the observed range of variation
of $x_{\rm max}$), and to a lesser extent, $x_{p1}$.
The bump strength $a_3$ is sensitive to $x_{02}$ and $x_{p1}$, whereas the
curvature $a_4$ is sensitive to $x_{02}$.
Table~\ref{t-isc} also shows that clustering, though still broadening
the UV bump compared to spheres, systematically shifts
the peak position of the bump to {\it smaller} wavenumbers,
which is contrary to what was found for single-Lorentz oscillator models
(and contrary to observations).
Thus, the inclusion of a background dielectric function
(arising from the oscillator associated with the FUV peak) inhibits the
conditions under which a shift to larger wavenumbers is possible.
Therefore, if the optical constants of Fig.~\ref{aoav_isc} are representative
of the interstellar UV feature carrier, then intrinsic variations in
chemical composition, rather than clustering, may be responsible for
this shift.

\section{Physical mechanisms of relevance}\label{Mechanisms}

In this section we discuss some of the laboratory studies examining
the UV properties of some carbonaceous materials of astrophysical relevance
and assess their usefulness in interpreting the results of the preceding
sections.

Fink et al.\ (1984) have derived $\epsilon_i$ of HAC thin films for
various annealing temperatures, $T_a$. As $T_a$ was increased,
the $\pi$--$\pi^*$
peak (associated with sp$^2$ bonding) shifted to smaller wavenumbers
and its strength increased. Equating a larger $T_a$ with less
hydrogenation, this means that we expect $x_{01}$ to shift to smaller
values and $x_{p1}$ to increase as the hydrogen content decreases.
Figures \ref{CoV_CCA} and \ref{CoV_cl} show qualitatively that trend.
On the other hand, as $T_a$ was increased, the $\sigma$--$\sigma^*$ peak
(observed for virtually all carbonaceous materials, including diamond)
initially at $\sim 10\mum1$, shifted to larger wavenumbers
($\sim 11\mum1$ at $T=540^{\circ}$C), but its strength stayed roughly the
same. Thus from this, one expects qualitatively less hydrogenated
material (larger $x_{p1}$) to exhibit larger $x_{02}$
(and thus, smaller curvature $a_4$). This appears to be contradicted by
the results of Fig.~\ref{aoav_isc}, though the comparison may not be
relevant since the widths observed in Fink et al.\ (1984) are much
broader than those considered here.

A dehydrogenation study of small HAC grains in {\it extinction}
has been carried out by Mennella et al.\ (1995a). The HAC
grains ``as produced'' show no UV bump, just a UV rise which is the
small $x$ tail of the $\sigma$--$\sigma^*$
electronic transition of carbon (observed at about $10.75\mum1$).
As the annealing temperature is increased from $250^{\circ}$C
to $800^{\circ}$C, a weak UV bump develops at around $5\mum1$
and gradually shifts to smaller $x$, gaining in strength. This
is compatible with the above trend observed in $\epsilon_i$.
When all hydrogen is lost (at around $800^{\circ}$C), the very broad
UV peak centered around $4\mum1$ resembles that of arc-evaporated
soot produced in an inert gas atmosphere (and UV features associated
with hydrogen-poor circumstellar environments; Blanco et al.\ 1995).
These features are much weaker and much broader than the interstellar
UV bump.

Simple dehydrogenation, by itself, cannot be the only process
responsible for the interstellar UV feature since it violates the
observed stability of the bump peak position. This
stability endures despite wide variations in temperature, density,
and UV flux arising from various interstellar environments.

Very recently, Mennella et al.\ (1996) have reported
a UV feature in small carbonaceous grains falling close
(at $2150~{\rm \AA}$) to the peak position of the interstellar UV bump. 
This feature was produced by subjecting small HAC grains to UV radiation
(corresponding to doses about 7 times less than typical
in the diffuse interstellar medium). More importantly, this feature
was shown to be {\it stable} in peak position when subjected to various
doses of UV radiation. However, the precise mechanism causing this
observed stability is still sketchy at the moment. After the UV
processing, the grains still contained a considerable amount of
hydrogen (about 0.3 relative to carbon by number, about half the
value found in the starting material). Unfortunately, the UV
features produced are considerably broader ($\ga 4~\mum1$)
than the interstellar UV feature. This could be due to the extreme
clustering observed for the grains and possibly size effects (since the
individual grains have a radius of about $50~{\rm \AA}$).
This laboratory cosmic dust analogue nevertheless looks extremely
promising in providing a better understanding of the optical properties
of the interstellar UV feature carrier.

\section{Conclusion}
In this paper, constraints on the possible clustering state and dielectric
function of the interstellar UV feature carrier were derived using various
models.

Compact clusters of graphite spheres were shown to satisfy qualitatively
the observational constraint that the UV bump peak position is very stable
while its width exhibits a wide range of values.
The main drawbacks of this model were that the peak position
was wrong, and that it did not allow for variations in chemical composition.
The latter is essential to account for the fact that variations in peak
position and width are not correlated, except for the widest bumps (observed
in dense interstellar regions where clustering of the grains is expected).

To remedy this, a single-Lorentz oscillator model in conjunction
with clustering was considered. It was shown to satisfy the
observational constraints, including the (correct) correlated shift in peak
position observed for very wide bumps.

But since the UV bump is usually assigned to some carbonaceous
material, then the carrier of this feature must also produce a FUV feature,
the tail of which would
be the FUV curvature observed along most lines of sight. A relatively simple
two-Lorentz oscillator model was shown to reproduce the extinction along
various lines of sight for a wide range of FUV curvatures.
However, in this model, clustering shifted the peak position to smaller
wavenumbers for the widest bumps, contrary to what is observed.
This model allowed a derivation of some optical constants of bump grains along
specific lines of sight, although the solution was not unique. Thus,
more measurements of the FUV part of the spectrum of carbonaceous interstellar
dust analogues would be most desirable to further constrain the model.

\begin{acknowledgements}
The computations in Sect.~\ref{Graphite} were performed on the CRAY T-90
of the Zentralinstitut f\"ur Angewandte Mathematik, in J\"ulich, Germany.
We acknowledge the constructive comments of the referee of this paper, 
B.T.\ Draine.
\end{acknowledgements}

\end{document}